\documentclass[letterpaper]{article} 
\usepackage{aaai24}  
\usepackage{times}  
\usepackage{helvet}  
\usepackage{courier}  
\usepackage[hyphens]{url}  
\usepackage{graphicx} 
\usepackage{natbib}  
\usepackage{caption} 
\usepackage{booktabs}
\usepackage{multirow}
\usepackage{pifont}
\usepackage{amsmath,amsthm,amssymb,amsfonts}

\usepackage{algorithm}
\usepackage{algorithmic}

\usepackage{newfloat}
\usepackage{listings}

\usepackage{newfloat}
\usepackage{listings}
\DeclareCaptionStyle{ruled}{labelfont=normalfont,labelsep=colon,strut=off} 
\floatstyle{ruled}
\newfloat{listing}{tb}{lst}{}
\floatname{listing}{Listing}
\title{CARE: A Large Scale CT Image Dataset and Clinical Applicable Benchmark Model for Rectal Cancer Segmentation}
\author {
	Hantao Zhang\textsuperscript{\rm 1,\rm 2}\equalcontrib,
	Weidong Guo\textsuperscript{\rm 1,\rm 2}\equalcontrib,
        Chenyang Qiu\textsuperscript{\rm 2,\rm 3},
        Shouhong Wan\textsuperscript{\rm 1,\rm 2}
        \footnote{Corresponding author.},
        Bingbing Zou\textsuperscript{\rm 2,\rm 3}\footnotemark[2],
        Wanqin Wang\textsuperscript{\rm 2,\rm 3}\footnotemark[2],
	Peiquan Jin\textsuperscript{\rm 1}
}
\affiliations {
	\textsuperscript{\rm 1} Key Laboratory of Electromagnetic Space Information, University of Science and Technology of China, Hefei, China
	\\
	\textsuperscript{\rm 2} Institute of Artificial Intelligence, Hefei Comprehensive National Science Center, Hefei, China
	\\
        \textsuperscript{\rm 3} The First Affiliated Hospital of Anhui Medical University

}

\begin{document}

\maketitle
\begin{abstract}
Rectal cancer segmentation of CT image plays a crucial role in timely clinical diagnosis, radiotherapy treatment, and follow-up. Although current segmentation methods have shown promise in delineating cancerous tissues, they still encounter challenges in achieving high segmentation precision. These obstacles arise from the intricate anatomical structures of the rectum and the difficulties in performing differential diagnosis of rectal cancer. Additionally, a major obstacle is the lack of a large-scale, finely annotated CT image dataset for rectal cancer segmentation. To address these issues, this work introduces a novel large scale rectal cancer CT image dataset CARE with pixel-level annotations for both normal and cancerous rectum, which serves as a valuable resource for algorithm research and clinical application development. Moreover, we propose a novel medical cancer lesion segmentation benchmark model named U-SAM. The model is specifically designed to tackle the challenges posed by the intricate anatomical structures of abdominal organs by incorporating prompt information. U-SAM contains three key components: promptable information (e.g., points) to aid in target area localization, a convolution module for capturing low-level lesion details, and skip-connections to preserve and recover spatial information during the encoding-decoding process. To evaluate the effectiveness of U-SAM, we systematically compare its performance with several popular segmentation methods on the CARE dataset. The generalization of the model is further verified on the WORD dataset. Extensive experiments demonstrate that the proposed U-SAM outperforms state-of-the-art methods on these two datasets. These experiments can serve as the baseline for future research and clinical application development.

\end{abstract}

\section{Introduction}

\begin{figure}
\begin{center}
\includegraphics[width=\linewidth,scale=1.00]{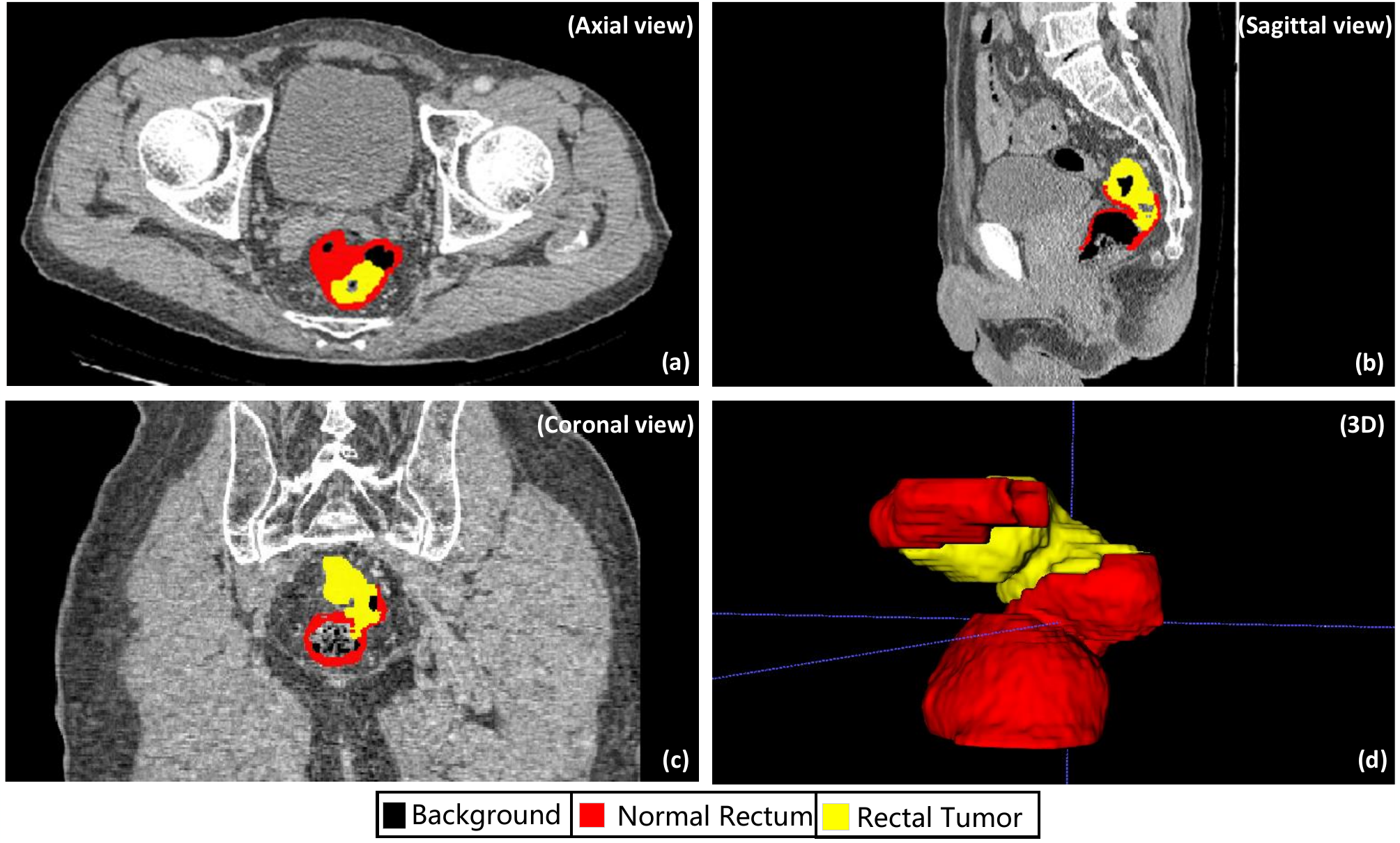}
\end{center}
   \caption{An example of the annotated normal rectum and rectal tumor in a CT scan. (a), (b), (c) denote the visualization in axial, sagittal, and coronal views, respectively. (d) represents the 3D rendering results of the annotations.}
\label{fig:dataset}
\end{figure}

Colorectal cancer ranks as the third most prevalent cancer worldwide and the second leading cause of cancer-related mortality. Notably, rectal cancer accounts for one-third of these cases \cite{keller2020multidisciplinary}. Timely detection and treatment can effectively halt the further deterioration of the patient's condition. Clinically, computed tomography (CT) is an integral part of rectal cancer's diagnostic workup, enabling comprehensive structural evaluation of the rectum. However, the process of oncologists delineating all colorectal cancer lesions from 3D volumes is both time-consuming and costly. Deep learning-based medical image segmentation has shown promise in reducing manual delineation efforts, but it necessitates a large-scale finely pixel-level annotated CT image dataset for effective training, particularly in diagnosing specific organ cancerous lesions. Unfortunately, there are currently no large-scale datasets available that cover rectal cancer with accurate and detailed annotations due to the extensive time and expertise required for such annotations.

\begin{figure*}
\begin{center}
\includegraphics[width=\linewidth,scale=1.00]{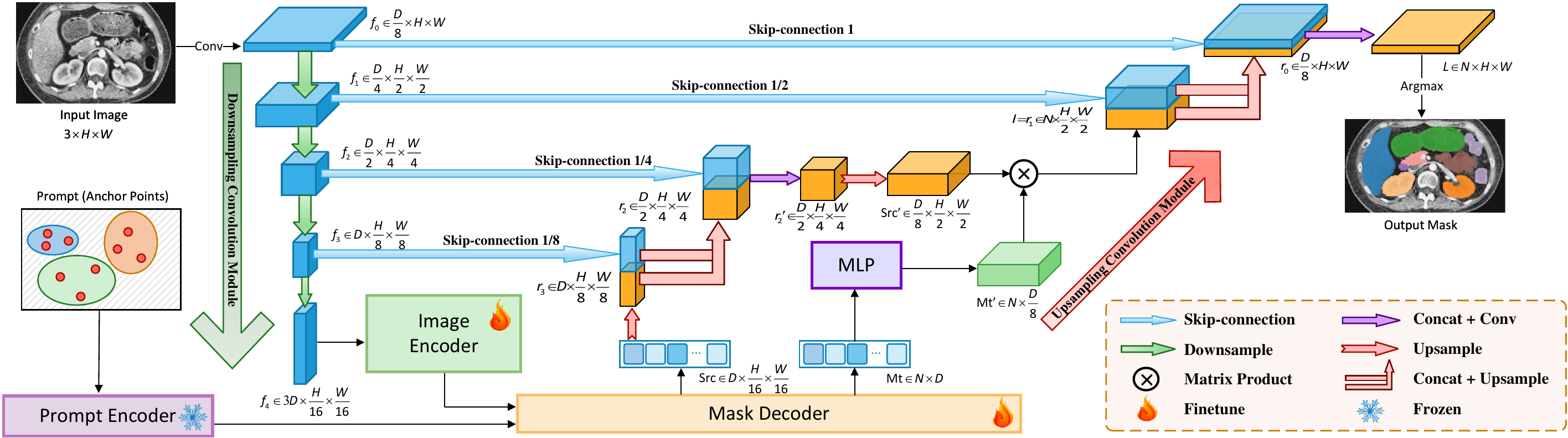}
\end{center}
   \caption{The framework of our proposed U-SAM.}
\label{fig:framework}
\end{figure*}

Previous studies \cite{landman2015miccai, valindria2018multi, luo2022word, ji2022amos, FLARE23, johnson2008accuracy} predominantly emphasized annotating organs rather than cancerous lesions. Although some datasets do include tumor annotations for certain cancer types, regrettably, there is a scarcity of annotated data specifically covering rectal cancer regions. In this study, we aim to address the gap in the field of rectal cancer segmentation by collecting a large-scale real clinical rectal cancer CT image dataset with careful pixel-level annotation. An example of image and annotation from the CARE (\textbf{C}linical \textbf{A}nnotation for \textbf{RE}ctal cancer segmentation) is illustrated in Fig. \ref{fig:dataset}. Collecting real clinical data poses significant challenges due to the difficulty in harmonizing source data with medical expertise and formatting. Moreover, annotating a large-scale medical image segmentation dataset, especially for rectal cancer is a costly and labor-intensive endeavor, necessitating much domain knowledge and clinical experience. In each case, a panel of experienced doctors, with more than 20 years of expertise, engage in thorough discussions to precisely identify the location and margins of the rectal cancer. Once a consensus is reached, detailed pixel-level annotations are meticulously crafted by one of them. On the whole, CARE is the first large-scale real clinical rectal cancer segmentation dataset from CT images.

Recently, there have been significant advancements in segmentation foundation models, particularly in the realm of natural image segmentation. Notably, the Segment anything model (SAM) \cite{kirillov2023segment} has played a vital role in revolutionizing natural image segmentation. Driven by the SAM's powerful generalization ability, several studies \cite{zhang2023survey, ma2023segment, wu2023medical, li2023polyp} have reported its performance in the medical domain. In the context of rectal cancer segmentation, achieving optimal results can be pretty challenging for conventional segmentation models(e.g., MissFormer\cite{huang2022missformer}, TransUnet \cite{chen2021transunet}, SwinUnet \cite{cao2022swin}, UCTransNet\cite{wang2022uctransnet}). This difficulty arises from the irregular shape of the rectum and the delicate nature of some intestinal walls, which possess thin wall thickness, as illustrated in Fig. \ref{fig:dataset}. These unique anatomical characteristics often pose hurdles to accurately segmenting the rectal region. Inspired by the success of SAM’s innovative promptable segmentation paradigm(e.g., bounding boxes, points, texts), we aim to introduce this novel paradigm to enhance the localization of the target rectum and address the intricate challenges posed by the complex nature of the rectal region. When the SAM model is applied in the medical field, it primarily encounters two challenges: 1. How to better capture the lesion details of cancerous organs and achieve effective transfer from natural scenes to medical scenes for SAM. 2. How to optimize the upsampling process of SAM's mask decoder to generate more accurate segmentation masks. 

To tackle these challenges, we propose a novel model named U-SAM(U-shaped SAM), as depicted in Fig. \ref{fig:framework}. Instead of solely relying on fine-tuning the SAM model \cite{ma2023segment,wu2023medical, li2023polyp}, we leverage SAM's promptable paradigm and integrate prior medical knowledge to develop a novel segmentation model. In U-SAM's image encoder, the Transformer encodes tokenized image patches from a convolutional neural network (CNN) feature map, serving as the input sequence to extract additional lesion details. As for U-SAM's decoder, it performs upsampling on the encoded features, which are then fused with high-resolution CNN feature maps to produce more precise segmentation masks.

The contributions of this paper can be summarized as: (1) We construct a large scale CT image dataset for rectal cancer segmentation. To the best of our knowledge, our dataset CARE is the first large scale CT image dataset with fine pixel-level annotations and scribble-based for the lesion information of rectal cancer. As shown in Fig. \ref{fig:dataset},  CARE obtains fine-grained annotations for both normal rectal regions and diseased tumor regions. Several state-of-the-art segmentation methods are evaluated on the CARE dataset. (2) Driven by the SAM's innovative promptable segmentation paradigm, we develop a novel prompting model named U-SAM for the medical domain. The model incorporates the convolution module designed to capture localized lesion information. Furthermore, we introduce an innovative variant of skip-connection to enhance the decoder's segmentation performance. (3) Experimental results demonstrate that our proposed U-SAM model obtains substantial improvements compared to existing methods and achieves state-of-the-art not just on the CARE but also on another abdominal organ CT image segmentation dataset WORD \cite{luo2022word}.
\section{Related Works}

\subsection{Abdominal Dataset}
Abdominal organ segmentation is essential in abdominal disease diagnosis, cancer treatment, and radiotherapy planning \cite{luo2022word}. Deep learning-based medical image segmentation has demonstrated the potential to reduce manual delineation efforts. However, it still faces challenges due to complex anatomical structures, indistinct boundaries of soft tissues, and significant variations in organ and tissue sizes. The vast disparity in knowledge requirements between natural and medical scenes makes it challenging for general models to achieve accurate abdominal segmentation, especially for cancerous regions. Therefore, specialized abdominal datasets are essential. Table \ref{table:dataset} provides an overview of the mainstream publicly available abdominal datasets, such as BTCV \cite{landman2015miccai}, CHAOS \cite{valindria2018multi}, WORD \cite{luo2022word}, and AMOS \cite{ji2022amos}, which focus on multiple organs in the abdomen. However, these datasets lack diagnostic and lesion information for the rectal region. For specific organ-focused datasets, such as Pancreas-CT\cite{FLARE23} and ACRIN 6664 \cite{johnson2008accuracy}, there is no corresponding cancer lesion information available. Although datasets like FLARE23\shortcite{FLARE23}, AutoPET\cite{gatidis2022whole}, LITS\cite{bilic2023liver}, and MSD\cite{antonelli2022medical} contain lesion information for some organs, they do not include relevant data for the rectum. To address this gap, we present a novel CT Rectal Cancer dataset in this work. All case samples are collected from real clinics and meticulously annotated by professional doctors. To the best of our knowledge, this is the largest cancer segmentation dataset, providing pixel-level annotations for both normal and cancerous rectal regions.

\subsection{SAM in Medicine}
Recently, the field of image segmentation has been revolutionized by the emergence of the large vision model Segment Anything Model (SAM)  \cite{kirillov2023segment} due to its generalization and versatility. With SAM's innovative promptable segmentation paradigm, a series of studies \cite{ma2023segment, wu2023medical, li2023polyp} have been conducted to explore the potential and limitations of SAM in medical image segmentation. The architecture of SAM consists of a ViT-based image encoder and a prompt-guided mask decoder. It was trained using a large dataset (SA-1B dataset). Mainstream SAM-based segmentation methods \cite{ma2023segment,wu2023medical, li2023polyp, zhang2023customized} primarily focus on transfer learning, which leverages the knowledge learned from large-scale natural image datasets to address specific challenges in the medical domain. These methods only modify a little architecture of the based SAM model. Some work\cite{ma2023segment, li2023polyp} explores how to freeze and fine-tune the modules in the SAM. While other methods \cite{wu2023medical, zhang2023customized} introduce adapter techniques to help the model better adapt to the downstream medical tasks. However, solely fine-tuning the SAM model does not always yield optimal results. As observed in \cite{wu2023medical}, SAM's overall performance still falls behind that of specialized models. Rather than solely concentrating on transferring learning knowledge from natural and medical images using the original SAM, our emphasis lies in designing an architecture specifically tailored to the medical field. Inspired by SAM's promptable paradigm, we integrate prior medical knowledge and devise a novel segmentation diagnostic model for medicine named U-SAM.

\section{The CARE Dataset}
\subsection{Overview}

The CARE rectal cancer dataset comprises CT scans from 399 patients, all diagnosed with rectal cancer. The original CT data and annotation files together exceed 74 GB in size. To ensure the imaging quality, all CT data has undergone enhancement. For a more accurate assessment of the tumor's condition, the patients providing the samples received corresponding contrast agents via blood vessel injection. This allows for easy identification of the rectal cancer area through changes in the contrast agent within the lesion and the mass enhancement. It is important to note that all CT images are anonymized. All clinical treatment details have been removed. The CARE dataset required around two years of meticulous collection, annotation, and rigorous review. This extensive effort has resulted in a valuable resource that bridges the gap in the field of rectal cancer segmentation.

\subsection{Professional data annotation}
The CARE dataset undergoes a rigorous annotation and review process to ensure its reliability. Initially, a gastrointestinal surgery clinician and a radiologist(both possessing over 20 years of experience) collaboratively analyze the patient's condition and the location of rectal cancer. Subsequently, more than ten gastrointestinal surgeons with extensive clinical expertise utilize ITK-SNAP\cite{yushkevich2006user} to delineate the diseased rectal area and the normal rectum slice-by-slice according to comprehensive information on different axial views. Finally, two oncology experts with over 20 years of experience thoroughly examine and revise these annotations. In cases of disagreement, they engage in discussions to reach the consensus of annotations, further enhancing the overall annotation quality.

\subsection{Data construction}
We conducted a random split of the CARE dataset into two subsets: 318 cases for training and 81 cases for testing. As the original CT data encompassed the entire body, we took the necessary steps to enhance training efficiency by eliminating irrelevant regions. Slices not containing the rectum were removed, and the corresponding images and labels were then packed into image-label pairs. In the end, we obtained 26,656 slice pairs for training and 6,424 pairs for testing.

\begin{table}[t]
\renewcommand\arraystretch{1.2} 
\scriptsize
\begin{center}
\begin{tabular}{c c c c c c}
\hline\hline
Dataset&Modality&Part&Pixel-level&Tumor&Number\\
\hline

BTCV \shortcite{landman2015miccai}&CT&Abdomen&\ding{51}&\ding{55}&50\\
CHAOS \shortcite{valindria2018multi}&CT\&MRI&Abdomen&\ding{51}&\ding{55}& 80\\
WORD \shortcite{luo2022word}&CT&Abdomen&\ding{51}&\ding{55}& 150\\
AMOS \shortcite{ji2022amos}&CT\&MRI&Abdomen&\ding{51}&\ding{55}& 600\\
Pancreas-CT\shortcite{FLARE23}&CT&Pancreas&\ding{51}&\ding{55}& 80 \\
ACRIN 6664\shortcite{johnson2008accuracy}& CT&Colon&\ding{55}&\ding{55}&825\\

\hline
FLARE23\shortcite{FLARE23}&CT&Abdomen&\ding{51}&\ding{51}& 500 \\
AutoPET\shortcite{gatidis2022whole}&CT\&PET&Body&\ding{51}&\ding{51}&900\\
LITS\shortcite{bilic2023liver}&CT&Liver&\ding{51}&\ding{51}& 200 \\
MSD\shortcite{antonelli2022medical}&CT&Colon&\ding{51}&\ding{51}&190\\
\textbf{CARE(ours)}&\textbf{CT}&\textbf{Rectum}&\ding{51}&\ding{51}&\textbf{399}\\

\hline\hline
\end{tabular}

\caption{Summary of several publicly available abdominal datasets. Modality: Medical data modalities. Part: Body parts covered by the dataset. Pixel-level: Whether the dataset contains pixel-level annotations. Tumor: Whether the dataset contains tumor information.}\label{table:dataset}
\end{center}

\end{table}

\section{Method}

\subsection{Overview}

Fig. \ref{fig:framework} illustrates the overview of our U-SAM framework. To the best of our knowledge, current SAM-based segmentation methods \cite{ma2023segment,wu2023medical, li2023polyp, zhang2023customized} mainly focus on transferring learning knowledge from natural and medical images without modifying too much architecture of the based SAM model. No efforts have been directed toward modifying the intrinsic architecture of SAM to align it more effectively with the requirements of the medical domain. In this study, we propose a novel architecture named U-SAM designed to enhance the segmentation capability within the medical domain. We integrate the promptable paradigm into U-SAM to enhance lesion localization and capture intricate details more effectively. Specifically, U-SAM contains three key components: promptable information (e.g., points) to aid in target area localization, a convolution module for capturing low-level lesion details, and skip-connections to preserve and recover spatial information during the encoding-decoding process. 
\subsection{Formulation}
 
In this section, we formulate the medical image segmentation task as follows:

Generally, given an image  $x \in \mathbb{R}^{C \times H \times W}$ and prompts $p$,  where $H \times W$ is the input resolution, and $C$ is the number of channels, the model predicts the pixel-wise mask $m \in \mathbb{R}^{H \times W}$ among all $N$ object classes including background. 

For the U-SAM, it only accepts anchor points as prompts. Thus, we define prompts as $p \in \mathbb{R}^{K \times 2}$, which consists of the 2-D coordinates of $K$ points. Note that the model allows null prompts as well. 

\subsection{U-SAM}
Firstly, we present the initial exploration stemming from the original SAM framework. It is important to note that the original SAM can exclusively produce foreground masks devoid of class labels. In contrast, our initial approach excels in generating pixel-wise label maps encompassing all $N$ object classes. In our initial approach, we adopt SAM's encoder-decoder and promptable paradigm. Specifically, we employ an image encoder and a prompt encoder to narrow the semantic gap between the input image and the provided prompts. These encoders intricately embed the input image and prompts into a shared latent space with D dimensions. Furthermore, we construct the query embedding $Q$ that encodes the segmentation task with prompts, which can be formulated as follows:
\begin{equation}
    \begin{gathered}
        pe = PromptEncoder(p),\\
        Q = QueryEncoder(pe)
    \end{gathered}
\end{equation}
where $pe$ indicates the embedding of prompts $p$ and $Q$ denotes the query embedding. 

Subsequently, we introduce SAM's mask decoder designed for mask generation. The decoder is based on the transformer architecture and takes image features and query embeddings $Q$ as inputs. It generates crucial raw mask information, forming the foundational elements for the subsequent mask-generation process. The raw mask information consists of mask source $Src \in \mathbb{R}^{D \times \frac{H}{16} \times \frac{W}{16}}$ and mask tokens $Mt \in \mathbb{R}^{N \times D}$. The mask tokens $Mt$ can be formulated as follows: 
\begin{equation}
     Mt = Concat(mt_1, mt_2, ..., mt_N)
\end{equation}
where $mt_i$ indicates the mask token of the $i^{th}$ class, and $N$ indicates the total number of object classes. 
Each mask token in $Mt$, containing the raw mask information of a unique object class, is later processed by the corresponding MLP module, which can be formulated as:
\begin{equation}
    \begin{gathered}
    mt'_i = MLP^i(mt_i),\\
    Mt' = Concat( mt'_1, mt'_2, ..., mt'_N)
    \end{gathered}
\end{equation}
where $MLP^i$ represents the MLP module belonging to the $i^{th}$ class, and  $mt'_i \in \mathbb{R}^{\frac{D}{8}}$  denotes the output mask token of the corresponding class with lower dimension of $\frac{D}{8}$. 
Additionally, we apply 4$\times$ upsampling to $Src$ and obtain $Src' \in \mathbb{R}^{\frac{D}{8} \times \frac{H}{4} \times \frac{W}{4}}$ with higher resolution, but fewer channels. 

After that, we obtained all the raw information. And then the low-resolution mask feature $l \in \mathbb{R}^{N \times {\frac{H}{4}} \times \frac{W}{4}}$ is generated as the following equation: 
\begin{equation}
    \label{eq_matrix_product}
    l = M’ \cdot Src’
\end{equation}
where ‘$\cdot$' represents matrix multiplication.
Furthermore, to restore the original resolution of $H \times W$, we upsample low-resolution feature $l$ with 4$\times$ scale bilinear interpolation to full-scale logits $L \in \mathbb{R}^{N \times H \times W}$. Finally, we generate the mask $m \in \mathbb{R}^{H \times W}$ by applying argmax to the $L$ on the first dimension.

As is elaborated above, following the framework of the original SAM, our initial method generally utilizes a two-step upsampling scheme to reconstruct the resolution. Instead of the aggressive long-stride upsampling strategy adopted in our initial exploration, we introduce a novel U-shaped architecture in the U-SAM. 

The framework of U-SAM is shown in Fig. \ref{fig:framework}. The pipeline of U-SAM can be generally divided into two major processes, including the downsampling encoder and upsampling decoder. In the following discussion, the image encoder is viewed as part of downsampling, and the mask decoder is defined as the beginning of upsampling.

\subsubsection{Downsampling Encoder}
In the downsampling encoder of our U-SAM, we extract feature representation using four consecutive downsampling blocks, each decreasing the resolution of the feature map by half. The process can be formulated as:
\begin{equation}
    f_{i+1} = Conv(MaxPool(f_i))
\end{equation}
where $f_i$ and $f_{i+1}$ indicate the input and output feature of the $i^{th}$ downsampling block, correspondingly. $MaxPool$ and  $Conv$ represents 2-D max pooling module and 2-D convolution module with a kernel size of 3, respectively. 

As is depicted in the left side of the Fig. \ref{fig:framework}, we obtain feature maps of all 5 layers, namely  $f_0 \in \mathbb{R}^{\frac{D}{8} \times H \times W}$, $f_1 \in \mathbb{R}^{\frac{D}{4} \times \frac{H}{2} \times \frac{W}{2}}$, $ f_2 \in \mathbb{R}^{\frac{D}{2} \times \frac{H}{4} \times \frac{W}{4}}$, $f_3 \in \mathbb{R}^{D \times \frac{H}{8} \times \frac{W}{8}}$ and $f_4 \in \mathbb{R}^{3D \times \frac{H}{16} \times \frac{W}{16}}$, where $D = 256$ represents the dimension of the SAM's latent space. Feature representation $f_4$ is then fed into the SAM's image decoder, while $f_0$, $f_1$, $f_2$ and $f_3$ later participate in the upsampling decoder process through skip-connections.

\subsubsection{Upsampling Decoder}
Given high-level feature representation $f_4$ and prompt embedding $pe$, the SAM's mask decoder generates raw mask information, i.e. mask source $Src \in \mathbb{R}^{D \times \frac{H}{16} \times \frac{W}{16}}$ and mask tokens $Mt \in \mathbb{R}^{N \times D}$. We obtain $Mt'$ the same way as our initial method. However, different from the initial implementation, U-SAM utilizes three consecutive 2$\times$ upsampling blocks, namely $UP^4$, $UP^3$ and $UP^2$ to reconstruct $Src$ into $Src' \in \mathbb{R}^{\frac{D}{8} \times \frac{H}{2} \times \frac{W}{2}}$. Corresponding upsampling operations can be formulated as:

\begin{equation}
    r_i = UP^{i+1}(r_{i+1}, f_{i+1})), \quad i=1,2,3
\end{equation}
where $r_{i+1}$ and $r_i$ correspond to the incoming and outgoing source features, respectively. $f_{i+1}$ denotes the image feature map  forwarding through skip-connection. 
Furthermore, in contrast to the initial approach that employs 4$\times$ bilinear interpolation to upsample low-resolution logits $l$, the recently introduced U-SAM incorporates an additional 2$\times$ upsampling block referred to as $UP^1$. This block is employed to restore complete-scale logits $L$, under the guidance of the skip-connected feature $f_0$.

\section{Experiments}

\begin{figure*}
\begin{center}
\includegraphics[width=\linewidth,scale=1.00]{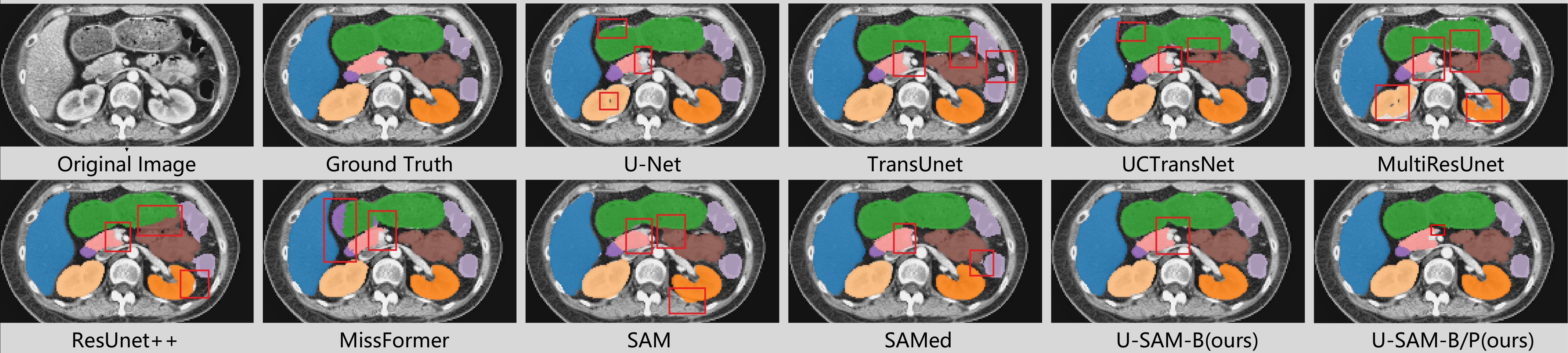}
\end{center}
   \caption{The qualitative comparison on the WORD dataset.}
\label{fig:word}
\end{figure*}

\begin{table}[t]
		\scriptsize
		\centering
		
		\begin{tabular}{@{} l @{\ \ }  l @{\ \ } l @{\ \ } l @{\ \ } l @{\ \ } l @{\ \ }@{\ \ \ } l@{}}
			\toprule
			\multirow{2}{*}{Method} & \multicolumn{2}{c}{Normal} & \multicolumn{2}{c}{Tumor} & \multicolumn{2}{c}{Mean} \\ \cmidrule(l){2-7} 
			    &Dice (\%)  & IoU (\%)           & Dice (\%)          & IoU (\%)  & Dice (\%)  & IoU (\%) \\ \midrule

			SwinUnet \shortcite{cao2022swin}&57.62  &40.46  &70.73  &54.72  &64.17 &47.59 \\
            V-Net\shortcite{milletari2016v}  &63.47 & 46.49 & 72.03 & 56.28 & 67.75 & 51.39 \\
			TransUnet \shortcite{chen2021transunet}  &60.21  &43.08  &70.69  &54.67  &65.45 &48.87 \\
            UCTransNet\shortcite{wang2022uctransnet}  &63.53  &46.55  &70.67  &54.64  &67.10 &50.59 \\
            AttenUnet\shortcite{oktay2018attention} &63.05  &46.04  &71.39  &55.50  &67.22 &50.77 \\
            ResUnet++\shortcite{jha2019resunet++}   &58.08  &40.93  &69.87  &53.69  &63.97 &47.31 \\
            MultiResUnet\shortcite{ibtehaz2020multiresunet} &62.25  &45.19  &72.11  &56.39  &67.18 &50.79 \\
            MissFormer\shortcite{huang2022missformer} &53.63  &36.64  &68.58  &52.19  &61.11 &44.41 \\

            \hline
                SAM-B\shortcite{kirillov2023segment} &60.95 &43.83 &71.00 &55.04 & 65.98 &49.44\\
                SAMed-H\shortcite{zhang2023customized} &60.05 &42.91 &70.72 &54.70 &65.38 &48.80\\
			\textbf{U-SAM-B(Ours)} &65.72 &48.94 &72.84 &57.28 &69.28 &53.11 \\
            \textbf{U-SAM-B/P(Ours)} &\textbf{70.27} &\textbf{54.17} &\textbf{76.49} &\textbf{61.94} &\textbf{73.38} &\textbf{58.05}
			\\
			\bottomrule
		\end{tabular}
		\caption{ Comparisons of performance with existing methods on the CARE Dataset. ‘B' denotes that the model utilizes the 'SAM-ViT-B', while ‘H' indicates the ‘SAM-ViT-H'.‘U-SAM-B/P' refers to the model incorporating 3 points prompt per class.}
		\label{table:care}
	\end{table}

\begin{table*}[t]
    \scriptsize
  \centering
    \setlength{\tabcolsep}{2.5pt}
    \resizebox{1\textwidth}{!}{
    \begin{tabular}{c|c|c|c|c|c|c|c|c|c|c|c|c|c|c|c|c|c}
\hline
    Model & Liver & Spleen &Kidney (L)& Kidney (R)& Stomach & Gallbladder & Esophagus & Pancreas  & Duodenum& Colon & Intestine & Adrenal & Rectum &Bladder  &HFL&HFR& Mean \\
\hline
    SwinUnet\shortcite{cao2022swin} & 94.62  & 92.72  & 89.32  & 89.31  & 88.85  & 70.92  & 66.36   & 70.11  & 53.20& 78.64  & 78.46  & 55.13  & 76.32  & 92.34  & 86.80  & 86.57  & 79.35\\
    V-Net\shortcite{milletari2016v} & 95.90 & 94.90 & 94.41 & 94.55 & 91.48 & 72.80 & 76.98 & 78.92 & 61.63& 82.06 & 83.30 & 67.65 & 78.54 & 94.51 & 88.46 & 88.72 & 84.05\\
    TransUnet\shortcite{chen2021transunet} & 95.46  & 93.21  & 91.47  & 91.63  & 90.01  & 70.99  & 70.61   & 75.38  & 55.47& 78.73  & 81.25  & 64.74  & 76.66  & 93.76  & 87.12  & 87.56  & 81.50\\
    UCTransNet\shortcite{wang2022uctransnet} & 95.19  & 94.18  & 94.27  & 94.62  & 89.04  & 65.83  & 68.67   & 73.30  & 58.44& 79.60  & 80.59  & 64.36  & 75.43  & 92.23  & 89.31  & 89.79  & 81.55\\
    FocalUnet\shortcite{naderi2022focal} & 93.21  & 89.54  & 88.64  & 88.68  & 83.43  & 61.29  & 57.83   & 60.57  & 45.20& 70.72  & 72.47  & 48.03  & 70.08  & 90.47  & 84.63  & 83.77  & 74.28\\
    R2Unet\shortcite{alom2019recurrent} & 84.73  & 90.35  & 90.56  & 87.78  & 80.21  & 59.56  & 71.12   & 72.64  & 49.74& 70.77  & 73.30  & 48.26  & 72.99  & 88.20  & 74.17  & 47.78  & 72.63\\
    ResUnet++\shortcite{jha2019resunet++} & 95.08  & 93.71  & 93.92  & 94.22  & 89.28  & 69.28  & 72.95   & 75.82  & 57.15& 79.80 & 80.73  & 65.59  & 75.27  & 93.20  & \textbf{92.26}  & \textbf{92.01}  & 82.52\\
    MultiResUnet\shortcite{ibtehaz2020multiresunet} & 95.19  & 93.73  & 93.12  & 93.33  & 90.73  & 69.83  & 73.11   & 75.33  & 60.36& 81.32 & 82.51  & 64.51  & 78.35  & 93.57  & 85.25  & 87.94  & 82.3\\
    MissFormer\shortcite{huang2022missformer} & 85.65 & 94.60 & 91.00 & 91.30 & 90.22 & 71.62 & 72.27 & 76.02 & 57.85& 80.44 & 80.87 & 64.02 & 76.55 & 93.53 & 87.26 & 86.90 & 81.89\\
    \hline
    
    SAM-B\shortcite{kirillov2023segment} & 94.50 &91.67 &89.44 & 88.91 &87.77 &59.83 &61.90 &70.15 &51.53 &71.91 &75.83 &51.71 &72.77 &91.91 &88.24 &88.34 &77.28  \\
    SAMed-H\shortcite{zhang2023customized}  &95.55 &94.77 &93.04 &93.13 &90.94 &73.29 &70.56 &73.73 &59.15&80.71 &81.09 &69.99 &78.62 &93.99 &87.74 &88.04 &82.27 \\
    \textbf{U-SAM-B(Ours)} &95.47	&\textbf{94.94} & \textbf{95.33} & \textbf{95.46} & 91.66 & \textbf{76.91} & 77.91 & 75.58 & \textbf{65.60} & \textbf{83.38} & 83.27 & 69.39 & 80.66 & 94.20 & 88.23 & 88.31 & 84.83 \\
    \textbf{U-SAM-B/P(Ours)} &\textbf{96.04} &94.85 &92.57 &92.68 &\textbf{91.96} &75.15 &\textbf{81.51} &\textbf{79.18} &65.54&82.73 &\textbf{83.99} &\textbf{73.15} &\textbf{82.94} &\textbf{95.03} &89.81 &89.86 &\textbf{85.44}  \\
\hline
    \end{tabular}%
    }

    \caption{Comparisons of performance with existing methods on the WORD Dataset. ‘B' denotes that the model utilizes the ‘SAM-ViT-B', while ‘H' indicates the ‘SAM-ViT-H'.‘U-SAM-B/P' refers to the model incorporating 3 points prompt per class. HFL:Head of Femur (L), HFR:Head of Femur(R).}
  \label{table:word}%
\end{table*}%

\subsection{Implementation Details}
In all experiments, we utilized PyTorch to implement our model, leveraging 8 NVIDIA 3090 GPU cards, each equipped with 24 GB of memory. To prevent overfitting, we applied two types of fundamental online data augmentations: random flipping and random rotating. It's worth noting that all of our experiments are based on the SAM-ViT-B. Additionally, we utilized the pre-trained weights of SAM \cite{kirillov2023segment} on natural images to expedite convergence and enhance training stability. Note that we froze the parameters in the prompt encoder of SAM during training for the same reason. Following the previous work \cite{cao2022swin, chen2021transunet}, we set the input resolution to 224 $\times$ 224. The batch size is used as 24. Our method is trained in an end-to-end manner, employing the Adam optimizer \cite{kingma2014adam}. To expedite faster convergence, the initial learning rates for the encoder and decoder parts are set to 0.001 and 0.0001, respectively. We also employ the combined cross entropy loss and dice loss as our loss function to train our network. For the CARE dataset, we utilize dice coefficient (Dice) and Intersection over Union (IoU) as the evaluation metrics, while for the WORD dataset, we report all organs' Dice.

\subsection{Comparison with the state-of-the-art methods}

\subsubsection{Comparison experiments on CARE}
To demonstrate the effectiveness of our proposed U-SAM, we compare our model with current state-of-the-art methods. We cover two types of methods for the comprehensive evaluation, including conventional segmentation methods(e.g., MissFormer\cite{huang2022missformer}, TransUnet \cite{chen2021transunet}, SwinUnet \cite{cao2022swin}, UCTransNet\cite{wang2022uctransnet}) and promptable paradigm model containing promptable information (e.g., SAM \cite{kirillov2023segment}, SAMed-H \cite{zhang2023customized}) Experimental results are reported in Table \ref{table:care}, where the best results are boldfaced.

Similar experimental results are obtained to the \cite{wu2023medical}, relying solely on fine-tuning the basic SAM model, falls behind the specialized segmentation models(e.g., UCTransNet\cite{wang2022uctransnet}). Moreover, while keeping the pre-trained weights locked, the approach of exclusively fine-tuning the adapter, as demonstrated in ‘SAMed-H'\cite{zhang2023customized}, also faces challenges in attaining ideal results. Rather than solely on transfer learning, we propose a novel U-shaped architecture to better adapt to the medical domain. Even without adding any promotable information, our proposed U-SAM outperforms all other competing methods, which validates the superiority of U-shaped architecture. Furthermore, thanks to the promotable paradigm, U-SAM obtains excellent results when adding 3 points promptable information per class. ‘U-SAM-B/P' achieves a 5.63\% mean Dice and 6.66\% mean IoU gain than the state-of-the-art methods.

\begin{figure}
\begin{center}
\includegraphics[width=\linewidth,scale=1.00]{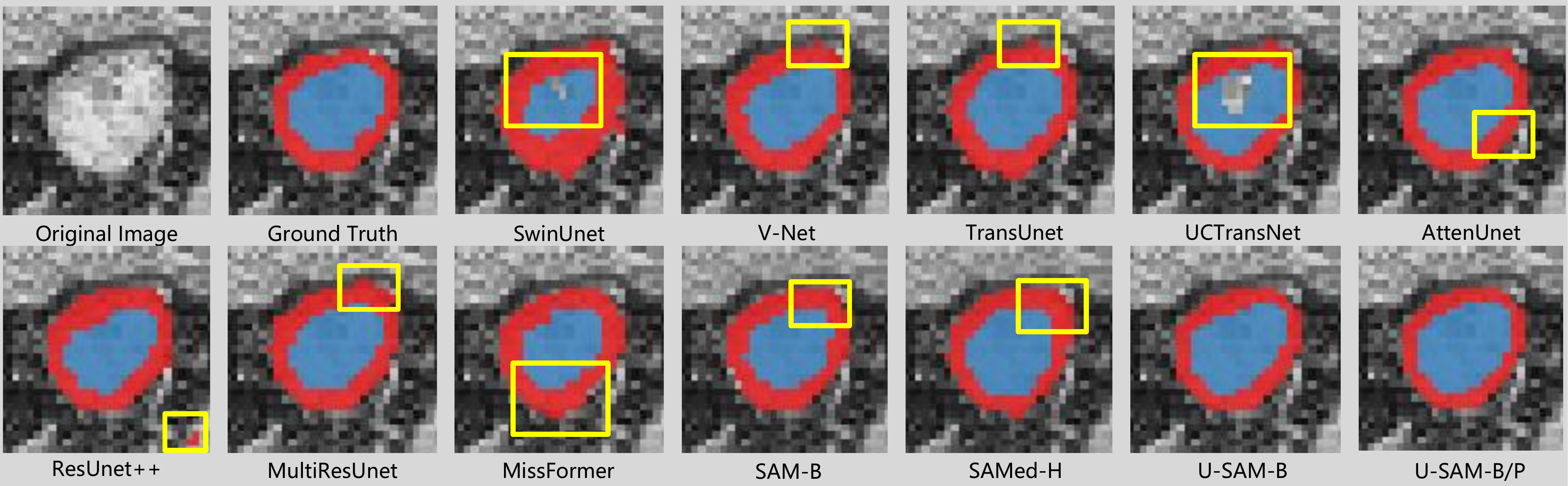}
\end{center}
   \caption{The qualitative comparison on the CARE dataset.}
\label{fig:care}
\vspace{-0.1in}
\end{figure}

\subsubsection{Generalization on WORD}
We further explore U-SAM's generalization on the WORD dataset. Experimental results are reported in Table \ref{table:word}, where the best results are boldfaced. The experiment results show that ‘U-SAM-B' has consistent improvements over prior arts, including specialized segmentation methods. Additionally, adding promptable information(3 points per class) not only achieves a faster convergence speed but also obtains better performance than the competing methods. These experiments again demonstrate U-SAM's excellent performance and reveal our model's remarkable generalization ability and versatility, which once again validates U-SAM's superiority of U-shaped architecture and promptable paradigm.

\subsection{Ablation Study}


\begin{table}[t]
		\scriptsize
		\centering
		
		\begin{tabular}{@{} l @{\ \ }  l @{\ \ } l @{\ \ } l @{\ \ } l @{\ \ } l @{\ \ }@{\ \ \ } l@{}}
			\toprule
			\multirow{2}{*}{Method} & \multicolumn{2}{c}{Normal} & \multicolumn{2}{c}{Tumor} & \multicolumn{2}{c}{Mean} \\ \cmidrule(l){2-7} 
			    &Dice (\%)            & IoU (\%)           & Dice (\%)          & IoU (\%)  & Dice (\%)          & IoU (\%) \\ \midrule
   
			S     &60.95 &43.83 &71.00 &55.04 & 65.98 &49.44 \\
			S+U &65.72 &48.94 &72.84 &57.28 &69.28 &53.11 \\
            S+U+1 point&67.65 &51.11 &73.94 &58.65 &70.79 &54.88 \\
            S+U+3 points&70.27 &54.17 &\textbf{76.49} &\textbf{61.94} &\textbf{73.38} &\textbf{58.05} \\
            S+U+5 points&\textbf{70.76} &\textbf{54.75} &75.62 &60.79 &73.19 &57.77 \\
			\bottomrule
		\end{tabular}
		\caption{Ablation experiments on the CARE dataset. ‘S' denotes the Baseline(SAM). ‘U' denotes the integration of convolutional modules and skip connections to form a U-shaped network.}

		\label{Table:ablation}
	\end{table}

\subsubsection{Ablation Studies on the Proposed Modules}

In this section, we conduct experiments to demonstrate U-SAM's superiority in U-shaped architecture and promptable (e.g., points) segmentation paradigm on the CARE dataset. As shown in Table \ref{Table:ablation}, the best results are boldfaced. Due to SAM's inherent structural limitations, it can not achieve ideal results in the medical domain. When adding the convolution module and linking the encoder and decoder by skip-connections to form the U-shaped model, it ahcieve a significant improvement of 3.3\% mean Dice and 3.67\% mean IoU.

Furthermore, we explore the potential of U-SAM' point promptable segmentation paradigm. Given only one anchor point prompt per class, 
‘S+U+1 point' ahcieves a improvement of 1.51\% mean Dice and 1.77\% mean IoU.When increasing the number of points to 3 per class, ‘S+U+3 points' achieves relatively best performance and gains a significant improvement of 4.10\% Mean Dice and 4.94\% Mean IoU. While the selection of promptable points involves random sampling, the increasement of points do not always lead to the better results. Hence, ‘S+U+5 points' fails to yield further improvements, leading us to adopt the utilization of three points in all subsequent optimal configurations.

\subsubsection{Ablation Studies on the Number of Skip-connections}

\begin{figure}
\begin{center}
\includegraphics[width=\linewidth,scale=1.00]{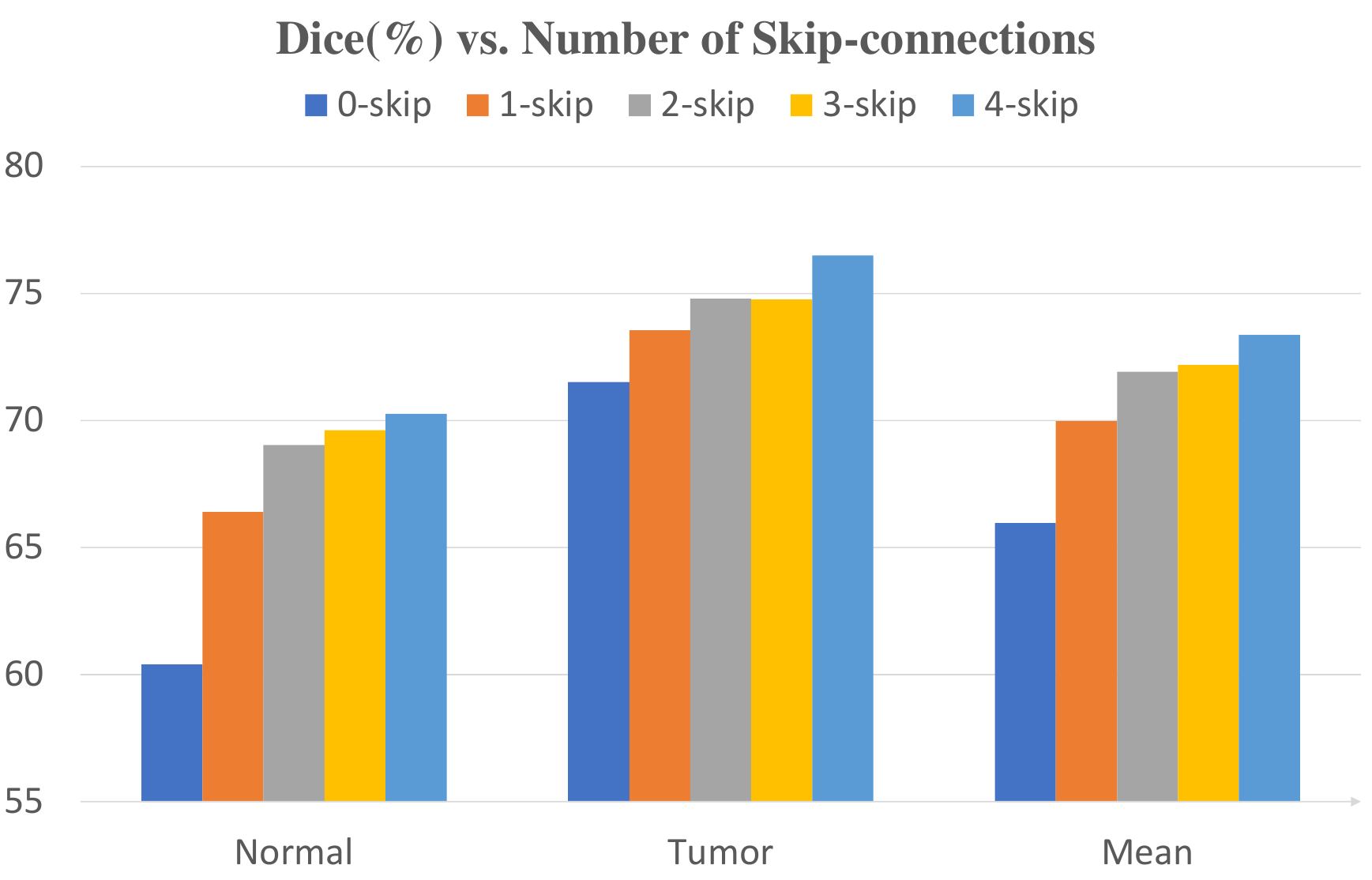}
\end{center}
   \caption{Ablation study results for Dice on the number of skip-connections in U-SAM.}
\label{fig:skip_dice}
\end{figure}

\begin{figure}
\begin{center}
\includegraphics[width=\linewidth,scale=1.00]{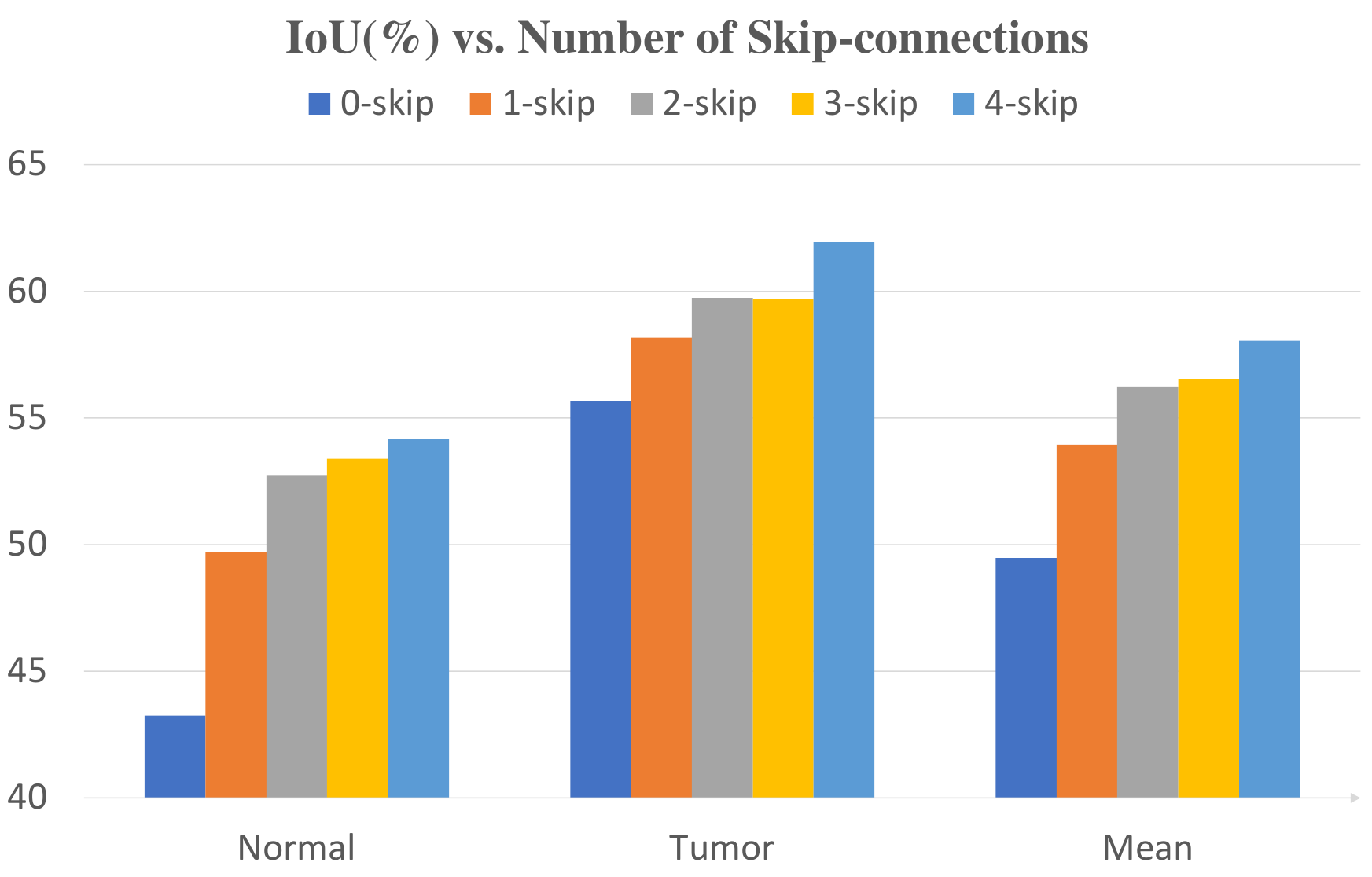}
\end{center}
   \caption{Ablation study results for IoU on the number of skip-connections in U-SAM.}
\label{fig:skip_iou}
\end{figure}

As previously explained, incorporating U-Net-like skip-connections within the U-SAM architecture plays a crucial role in retaining and restoring spatial information throughout the encoding-decoding process. Consequently, we conduct an ablation study concerning the number of skip-connections to ascertain the optimal configuration for these connections. All experiments in this section utilize 3  points per class. Specifically, we vary the skip-connections from 0 to 4 and carry the experiments on the CARE dataset. We manipulate the number of skip-connections, ranging from 0 to 4, and conduct the experiments on the CARE dataset. It should be noticed that ‘0-skip' uses no skip-connections, ‘1-skip' integrates a sole skip-connection at a scale of $\frac{1}{8}$, while ‘2-skip' implements two skip-connections, strategically positioned at both $\frac{1}{8}$ and $\frac{1}{4}$ scales, and so forth.

The experiment results for Dice and IoU are visually represented in Fig. \ref{fig:skip_dice} and Fig. \ref{fig:skip_iou}, correspondingly. The results show that adding more skip-connections generally leads to a better segmentation performance. The best mean Dice and mean IoU are achieved by inserting skip-connections to all four intermediate upsampling steps of $\frac{1}{8}$, $\frac{1}{4}$, $\frac{1}{2}$ and $1$ resolution scales. Thus, this best configuration is adopted in our U-SAM. It is also worth mentioning that significant enhancements have been attained for both normal rectum and tumor. This indicates the effectiveness of integrating U-Net-like skip-connections within the SAM framework, which benefit the extraction of intricate details, such as the precise delineation of the rectum's borders.

\subsubsection{Visualizations}
We visualize the segmentation results of the comparable models in Fig. \ref{fig:care} and Fig. \ref{fig:word}. Notably, the boxes highlight the regions where the model makes mistakes compared to the ground truth. It shows that our U-SAM generates better segmentation results, which are more similar to the ground truth than the baseline model results.

It shows that our proposed method excels in accentuating salient areas while eliminating perplexing false positive lesions and producing coherent boundaries. These insights imply that U-SAM can achieve more refined segmentation while preserving intricate shape information, particularly along the borders of the normal rectum and tumor regions, as illustrated in Fig. \ref{fig:care}.

\section{Conclusion}
In this paper, we construct the first large scale CT rectal cancer dataset CARE with pixel-level annotations for both normal and cancerous rectum, effectively addressing gaps within the realm of rectal cancer segmentation. Inspired by the success of SAM’s innovative promptable (e.g., points) segmentation paradigm, we develop a novel model U-SAM to achieve better rectal cancer segmentation. The U-SAM model adopts a U-shaped architecture, rectifying the inherent structural limitations of SAM when applied to medical image analysis. It integrates a convolution module to extract localized information more effectively while employing skip-connections to both preserve and restore spatial information throughout the encoding-decoding process. Extensive experiments demonstrate that the proposed U-SAM outperforms state-of-the-art methods on CARE and WORD datasets. In the future, we will still work on extending the CARE dataset to be more extensive and further explore the potential of U-SAM within the domain of medical image segmentation.

\bibliography{aaai24}

\begin{thebibliography}{28}
\providecommand{\natexlab}[1]{#1}

\bibitem[{FLA(2023)}]{FLARE23}
 2023.
\newblock MICCAI FLARE23: Fast, Low-resource, and Accurate oRgan and Pan-cancer
  sEgmentation in Abdomen CT.
\newblock \url{https://codalab.lisn.upsaclay.fr/competitions/12239}.

\bibitem[{Alom et~al.(2019)Alom, Yakopcic, Hasan, Taha, and
  Asari}]{alom2019recurrent}
Alom, M.~Z.; Yakopcic, C.; Hasan, M.; Taha, T.~M.; and Asari, V.~K. 2019.
\newblock Recurrent residual U-Net for medical image segmentation.
\newblock \emph{Journal of Medical Imaging}, 6(1): 014006--014006.

\bibitem[{Antonelli et~al.(2022)Antonelli, Reinke, Bakas, Farahani,
  Kopp-Schneider, Landman, Litjens, Menze, Ronneberger, Summers
  et~al.}]{antonelli2022medical}
Antonelli, M.; Reinke, A.; Bakas, S.; Farahani, K.; Kopp-Schneider, A.;
  Landman, B.~A.; Litjens, G.; Menze, B.; Ronneberger, O.; Summers, R.~M.;
  et~al. 2022.
\newblock The medical segmentation decathlon.
\newblock \emph{Nature communications}, 13(1): 4128.

\bibitem[{Bilic et~al.(2023)Bilic, Christ, Li, Vorontsov, Ben-Cohen, Kaissis,
  Szeskin, Jacobs, Mamani, Chartrand et~al.}]{bilic2023liver}
Bilic, P.; Christ, P.; Li, H.~B.; Vorontsov, E.; Ben-Cohen, A.; Kaissis, G.;
  Szeskin, A.; Jacobs, C.; Mamani, G. E.~H.; Chartrand, G.; et~al. 2023.
\newblock The liver tumor segmentation benchmark (lits).
\newblock \emph{Medical Image Analysis}, 84: 102680.

\bibitem[{Cao et~al.(2022)Cao, Wang, Chen, Jiang, Zhang, Tian, and
  Wang}]{cao2022swin}
Cao, H.; Wang, Y.; Chen, J.; Jiang, D.; Zhang, X.; Tian, Q.; and Wang, M. 2022.
\newblock Swin-unet: Unet-like pure transformer for medical image segmentation.
\newblock In \emph{European conference on computer vision}, 205--218. Springer.

\bibitem[{Chen et~al.(2021)Chen, Lu, Yu, Luo, Adeli, Wang, Lu, Yuille, and
  Zhou}]{chen2021transunet}
Chen, J.; Lu, Y.; Yu, Q.; Luo, X.; Adeli, E.; Wang, Y.; Lu, L.; Yuille, A.~L.;
  and Zhou, Y. 2021.
\newblock Transunet: Transformers make strong encoders for medical image
  segmentation.
\newblock \emph{arXiv preprint arXiv:2102.04306}.

\bibitem[{Gatidis et~al.(2022)Gatidis, Hepp, Fr{\"u}h, La~Foug{\`e}re,
  Nikolaou, Pfannenberg, Sch{\"o}lkopf, K{\"u}stner, Cyran, and
  Rubin}]{gatidis2022whole}
Gatidis, S.; Hepp, T.; Fr{\"u}h, M.; La~Foug{\`e}re, C.; Nikolaou, K.;
  Pfannenberg, C.; Sch{\"o}lkopf, B.; K{\"u}stner, T.; Cyran, C.; and Rubin, D.
  2022.
\newblock A whole-body FDG-PET/CT Dataset with manually annotated Tumor
  Lesions.
\newblock \emph{Scientific Data}, 9(1): 601.

\bibitem[{Huang et~al.(2022)Huang, Deng, Li, Yuan, and
  Fu}]{huang2022missformer}
Huang, X.; Deng, Z.; Li, D.; Yuan, X.; and Fu, Y. 2022.
\newblock Missformer: An effective transformer for 2d medical image
  segmentation.
\newblock \emph{IEEE Transactions on Medical Imaging}.

\bibitem[{Ibtehaz and Rahman(2020)}]{ibtehaz2020multiresunet}
Ibtehaz, N.; and Rahman, M.~S. 2020.
\newblock MultiResUNet: Rethinking the U-Net architecture for multimodal
  biomedical image segmentation.
\newblock \emph{Neural networks}, 121: 74--87.

\bibitem[{Jha et~al.(2019)Jha, Smedsrud, Riegler, Johansen, De~Lange,
  Halvorsen, and Johansen}]{jha2019resunet++}
Jha, D.; Smedsrud, P.~H.; Riegler, M.~A.; Johansen, D.; De~Lange, T.;
  Halvorsen, P.; and Johansen, H.~D. 2019.
\newblock Resunet++: An advanced architecture for medical image segmentation.
\newblock In \emph{2019 IEEE international symposium on multimedia (ISM)},
  225--2255. IEEE.

\bibitem[{Ji et~al.(2022)Ji, Bai, Ge, Yang, Zhu, Zhang, Li, Zhanng, Ma, Wan
  et~al.}]{ji2022amos}
Ji, Y.; Bai, H.; Ge, C.; Yang, J.; Zhu, Y.; Zhang, R.; Li, Z.; Zhanng, L.; Ma,
  W.; Wan, X.; et~al. 2022.
\newblock Amos: A large-scale abdominal multi-organ benchmark for versatile
  medical image segmentation.
\newblock \emph{Advances in Neural Information Processing Systems}, 35:
  36722--36732.

\bibitem[{Johnson et~al.(2008)Johnson, Chen, Toledano, Heiken, Dachman, Kuo,
  Menias, Siewert, Cheema, Obregon et~al.}]{johnson2008accuracy}
Johnson, C.~D.; Chen, M.-H.; Toledano, A.~Y.; Heiken, J.~P.; Dachman, A.; Kuo,
  M.~D.; Menias, C.~O.; Siewert, B.; Cheema, J.~I.; Obregon, R.~G.; et~al.
  2008.
\newblock Accuracy of CT colonography for detection of large adenomas and
  cancers.
\newblock \emph{New England Journal of Medicine}, 359(12): 1207--1217.

\bibitem[{Keller et~al.(2020)Keller, Berho, Perez, Wexner, and
  Chand}]{keller2020multidisciplinary}
Keller, D.~S.; Berho, M.; Perez, R.~O.; Wexner, S.~D.; and Chand, M. 2020.
\newblock The multidisciplinary management of rectal cancer.
\newblock \emph{Nature Reviews Gastroenterology \& Hepatology}, 17(7):
  414--429.

\bibitem[{Kingma and Ba(2014)}]{kingma2014adam}
Kingma, D.~P.; and Ba, J. 2014.
\newblock Adam: A method for stochastic optimization.
\newblock \emph{arXiv preprint arXiv:1412.6980}.

\bibitem[{Kirillov et~al.(2023)Kirillov, Mintun, Ravi, Mao, Rolland, Gustafson,
  Xiao, Whitehead, Berg, Lo et~al.}]{kirillov2023segment}
Kirillov, A.; Mintun, E.; Ravi, N.; Mao, H.; Rolland, C.; Gustafson, L.; Xiao,
  T.; Whitehead, S.; Berg, A.~C.; Lo, W.-Y.; et~al. 2023.
\newblock Segment anything.
\newblock \emph{arXiv preprint arXiv:2304.02643}.

\bibitem[{Landman et~al.(2015)}]{landman2015miccai}
Landman, B.; et~al. 2015.
\newblock MICCAI multi-atlas labeling beyond the cranial vault--workshop and
  challenge.
\newblock In \emph{Proc. MICCAI Multi-Atlas Labeling Beyond Cranial
  Vault—Workshop Challenge}, volume~5, 12.

\bibitem[{Li, Hu, and Yang(2023)}]{li2023polyp}
Li, Y.; Hu, M.; and Yang, X. 2023.
\newblock Polyp-sam: Transfer sam for polyp segmentation.
\newblock \emph{arXiv preprint arXiv:2305.00293}.

\bibitem[{Luo et~al.(2022)Luo, Liao, Xiao, Chen, Song, Zhang, Li, Metaxas,
  Wang, and Zhang}]{luo2022word}
Luo, X.; Liao, W.; Xiao, J.; Chen, J.; Song, T.; Zhang, X.; Li, K.; Metaxas,
  D.~N.; Wang, G.; and Zhang, S. 2022.
\newblock WORD: A large scale dataset, benchmark and clinical applicable study
  for abdominal organ segmentation from CT image.
\newblock \emph{Medical Image Analysis}, 82: 102642.

\bibitem[{Ma and Wang(2023)}]{ma2023segment}
Ma, J.; and Wang, B. 2023.
\newblock Segment anything in medical images.
\newblock \emph{arXiv preprint arXiv:2304.12306}.

\bibitem[{Milletari, Navab, and Ahmadi(2016)}]{milletari2016v}
Milletari, F.; Navab, N.; and Ahmadi, S.-A. 2016.
\newblock V-net: Fully convolutional neural networks for volumetric medical
  image segmentation.
\newblock In \emph{2016 fourth international conference on 3D vision (3DV)},
  565--571. Ieee.

\bibitem[{Naderi et~al.(2022)Naderi, Givkashi, Piri, Karimi, and
  Samavi}]{naderi2022focal}
Naderi, M.; Givkashi, M.; Piri, F.; Karimi, N.; and Samavi, S. 2022.
\newblock Focal-UNet: UNet-like Focal Modulation for Medical Image
  Segmentation.
\newblock \emph{arXiv preprint arXiv:2212.09263}.

\bibitem[{Oktay et~al.(2018)Oktay, Schlemper, Folgoc, Lee, Heinrich, Misawa,
  Mori, McDonagh, Hammerla, Kainz et~al.}]{oktay2018attention}
Oktay, O.; Schlemper, J.; Folgoc, L.~L.; Lee, M.; Heinrich, M.; Misawa, K.;
  Mori, K.; McDonagh, S.; Hammerla, N.~Y.; Kainz, B.; et~al. 2018.
\newblock Attention u-net: Learning where to look for the pancreas.
\newblock \emph{arXiv preprint arXiv:1804.03999}.

\bibitem[{Valindria et~al.(2018)Valindria, Pawlowski, Rajchl, Lavdas, Aboagye,
  Rockall, Rueckert, and Glocker}]{valindria2018multi}
Valindria, V.~V.; Pawlowski, N.; Rajchl, M.; Lavdas, I.; Aboagye, E.~O.;
  Rockall, A.~G.; Rueckert, D.; and Glocker, B. 2018.
\newblock Multi-modal learning from unpaired images: Application to multi-organ
  segmentation in CT and MRI.
\newblock In \emph{2018 IEEE winter conference on applications of computer
  vision (WACV)}, 547--556. IEEE.

\bibitem[{Wang et~al.(2022)Wang, Cao, Wang, and Zaiane}]{wang2022uctransnet}
Wang, H.; Cao, P.; Wang, J.; and Zaiane, O.~R. 2022.
\newblock Uctransnet: rethinking the skip connections in u-net from a
  channel-wise perspective with transformer.
\newblock In \emph{Proceedings of the AAAI conference on artificial
  intelligence}, volume~36, 2441--2449.

\bibitem[{Wu et~al.(2023)Wu, Fu, Fang, Liu, Wang, Xu, Jin, and
  Arbel}]{wu2023medical}
Wu, J.; Fu, R.; Fang, H.; Liu, Y.; Wang, Z.; Xu, Y.; Jin, Y.; and Arbel, T.
  2023.
\newblock Medical SAM Adapter: Adapting Segment Anything Model for Medical
  Image Segmentation.
\newblock \emph{arXiv preprint arXiv:2304.12620}.

\bibitem[{Yushkevich et~al.(2006)Yushkevich, Piven, Hazlett, Smith, Ho, Gee,
  and Gerig}]{yushkevich2006user}
Yushkevich, P.~A.; Piven, J.; Hazlett, H.~C.; Smith, R.~G.; Ho, S.; Gee, J.~C.;
  and Gerig, G. 2006.
\newblock User-guided 3D active contour segmentation of anatomical structures:
  significantly improved efficiency and reliability.
\newblock \emph{Neuroimage}, 31(3): 1116--1128.

\bibitem[{Zhang et~al.(2023)Zhang, Zheng, Li, Qiao, Kang, Shan, Zhang, Qin,
  Rameau, Bae et~al.}]{zhang2023survey}
Zhang, C.; Zheng, S.; Li, C.; Qiao, Y.; Kang, T.; Shan, X.; Zhang, C.; Qin, C.;
  Rameau, F.; Bae, S.-H.; et~al. 2023.
\newblock A Survey on Segment Anything Model (SAM): Vision Foundation Model
  Meets Prompt Engineering.
\newblock \emph{arXiv preprint arXiv:2306.06211}.

\bibitem[{Zhang and Liu(2023)}]{zhang2023customized}
Zhang, K.; and Liu, D. 2023.
\newblock Customized segment anything model for medical image segmentation.
\newblock \emph{arXiv preprint arXiv:2304.13785}.

\end{thebibliography}

\end{document}